\documentclass[conference]{IEEEtran}
\IEEEoverridecommandlockouts
\usepackage{cite}
\usepackage{amsmath,amssymb,amsfonts}
\usepackage{algorithmic}
\usepackage{graphicx}
\usepackage{textcomp}
\usepackage{xcolor}

\usepackage{algorithm}
\usepackage{multirow}

\usepackage{booktabs}

\def\BibTeX{{\rm B\kern-.05em{\sc i\kern-.025em b}\kern-.08em
    T\kern-.1667em\lower.7ex\hbox{E}\kern-.125emX}}
\begin{document}

\title{Semantic-aware Transmission for Robust Point Cloud Classification 
}

\author{\IEEEauthorblockN{Tianxiao Han, Kaiyi Chi, Qianqian Yang, Zhiguo Shi}
\IEEEauthorblockA{ College of Information Science and Electronic Engineering, Zhejiang University, Hangzhou 310007, China \\
Email: 
\{txhan, kaiyichi, qianqianyang20, shizg\}@zju.edu.cn
}

}

\maketitle

\begin{abstract}
As three-dimensional (3D) data acquisition devices become increasingly prevalent, the demand for 3D point cloud transmission is growing. In this study, we introduce a semantic-aware communication system for robust point cloud classification that capitalizes on the advantages of pre-trained Point-BERT models. Our proposed method comprises four main components: the semantic encoder, channel encoder, channel decoder, and semantic decoder. By employing a two-stage training strategy, our system facilitates efficient and adaptable learning tailored to the specific classification tasks. The results show that the proposed system achieves classification accuracy of over 89\% when SNR is higher than 10 dB and still maintains accuracy above 66.6\% even at SNR of 4 dB. Compared to the existing method, our approach performs at 0.8\% to 48\%  better across different SNR values, demonstrating robustness to channel noise. Our system also achieves a balance between accuracy and speed, being computationally efficient while maintaining high classification performance under noisy channel conditions. This adaptable and resilient approach holds considerable promise for a wide array of 3D scene understanding applications, effectively addressing the challenges posed by channel noise.

\end{abstract}


\section{Introduction}

With the development of three-dimensional (3D) acquisition devices, such as LiDARs, and RGB-D cameras, 3D vision has recently become a prominent research topic \cite{Stereo}. Among various types of 3D data representations, point cloud is a widely used data format due to its effective presenting ability. However, transmitting point cloud data is challenging since the data size is relatively large.  For instance, a 3D point cloud generated by LiDAR with 0.7 million points requires a bandwidth of around 500 Mbps with 30 fps \cite{data}, which is extremely high for standard communication systems.

To address the issue, there are some works proposing effective compression methods before the transmission to reduce the communication overhead, such as MPEG's G-PCC \cite{g-pcc} and Google's Draco \cite{draco}. Although these methods have made strides in improving transmission efficiency, G-PCC and Draco have their advantages but suffer from some drawbacks \cite{survey}, such as being susceptible to noise and lacking effective acceleration on GPUs.

The authors in \cite{segeme} proposed a point compression algorithm for LiDAR point clouds to achieve quasi real-time transmissions in autonomous driving scenarios. The authors utilized a semantic segmentation network to identify the more valuable data and then compressed the resulting point cloud with Draco. The work in \cite{aitransformer} proposed a method for efficient point cloud compression by leveraging the inter-frame redundancy and spatial correlation between point clouds. However, the proposed method lacks robustness against channel noise.

In fact, it is not always necessary to reconstruct the 3D objects at the receiver side. For example, if the receiver is only interested in the classification of the object, the transmitter can transmit information that contributes most to the classification results, which is also called semantic communication. Different from the traditional communication systems, semantic communication extracts and transmits task-relevant information and filters out task-irrelevant information \cite{siqi}, which can significantly reduce the communication overhead. So far, researchers have proposed numerous semantic communication models on text transmission \cite{deepsc}, image transmission \cite{multi-level} and speech transmission\cite{speech}, and the results have demonstrated the superiority over traditional communication methods.  However, the irregular, unstructured and unordered nature of point clouds makes feature extraction challenging and deep learning methods more susceptible to noise. The authors in \cite{Branchy} proposed a graph neural network-based framework for efficient point cloud processing using edge computing platforms, they utilized a learning-based semantic communication network \cite{jscc} for the intermediate feature compression, which can significantly reduce the communication overhead. But under bad channel conditions, such as SNR below 16dB as shown in their paper, the accuracy drops sharply, for example, the accuracy drops 17\% from 10dB to 12dB and 19\% from 10dB to 8dB.

In this paper, we propose a task-oriented semantic communication system for effectively and robustly transmitting point cloud data. We leverage farthest point sampling (FPS) \cite{fast} to extract key points, then utilize k-nearest neighbors (kNN) \cite{knn} to generate corresponding sub-clouds. Subsequently, the sub-clouds are processed by a semantic encoder, which employs the pre-trained transformer-based model, Point-BERT \cite{point-bert}, to efficiently extract robust semantic information that is relevant to specific classification tasks. The resulting task-relevant feature vector is transmitted via a physical channel and decoded at the receiver side using a semantic decoder. The simulation results demonstrate the effectiveness and robustness of our proposed system in stably transmitting point cloud data.

The remainder of the paper is organized as follows: Section \uppercase\expandafter{\romannumeral2} presents the system model. Section \uppercase\expandafter{\romannumeral3} describes the proposed model in detail.
Section \uppercase\expandafter{\romannumeral4} presents the simulation results to validate the effectiveness of the proposed model. Section \uppercase\expandafter{\romannumeral5} concludes the paper.

\begin{figure}[htbp]
\centering
\includegraphics[width=\linewidth]{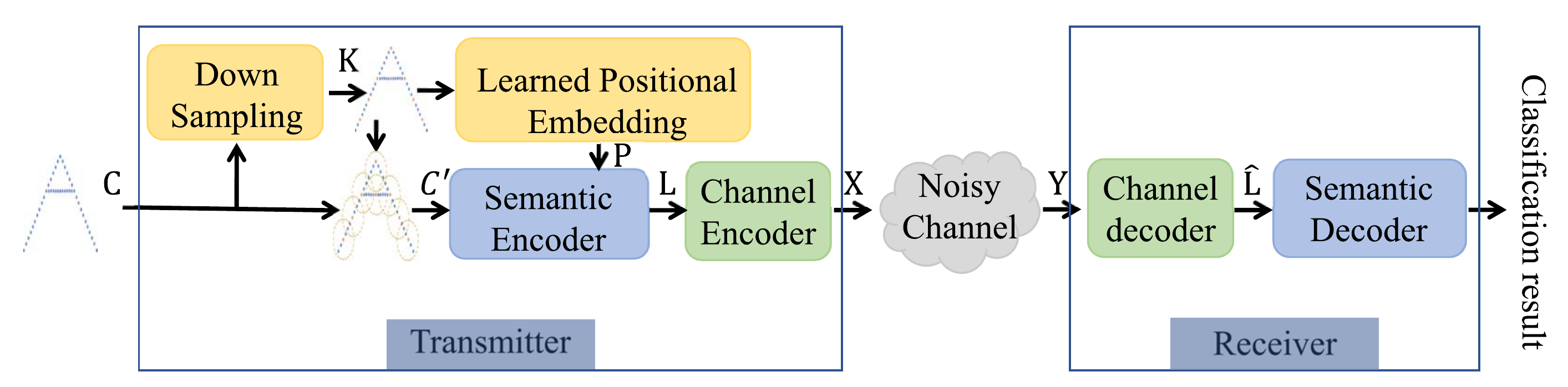}
\caption{The system model of the proposed point cloud communication system}
\label{fig:framework}
\end{figure}

\section{System Model}

In this section, we provide an overview of the system model for the considered semantic communication system for point cloud classification, which aims to mitigate the impact of noise on the semantic network of the point cloud. Moreover, we introduce the performance metrics used to evaluate our proposed approach.

\subsection{Transmitter and Receiver}

As shown in Fig.~\ref{fig:framework}, the proposed semantic communication system for point cloud classification consists of two main components: a transmitter network and a receiver network. The transmitter network receives a dense point cloud $\boldsymbol C$ as input. At first, the transmitter applies FPS to downsample the input point cloud, obtaining approximately 0.8\% of the key points $\boldsymbol K$ as the crucial basis for transmission of semantic information. Then those key points are fed into a learned positional embedding layer to generate an embedding vector for each key points to represent the geometry information. Meanwhile, the transmitter performs kNN on the input dense point cloud and the key points, generating sub-clouds of neighboring points for each key point. Subsequently, both the embedding vector and the sub-clouds are fed into a semantic encoder to obtain the latent representation $\boldsymbol L$ for transmission. Then the latent representation serves as the input to a channel encoder for encoding and transmission. The channel encoder maps $\boldsymbol L$ into symbol vector $\boldsymbol X$, which is transmitted via the physical channel.

At the receiver end, the received signal is given by:

\begin{equation}
\boldsymbol Y = \boldsymbol {\mathbf{h}} \ast \boldsymbol X + \boldsymbol w,
\label{channel}
\end{equation}
where $\boldsymbol h$ represents the channel coefficients, and $\boldsymbol{w}$ denotes the noise that follows the complex Gaussian distribution $\mathcal{CN}(0, \sigma^2\mathbf I)$, and $\sigma^2$ is the noise variance.

The receiver consists of a channel decoder, and a semantic decoder. 
The channel decoder decodes the semantic vectors transmitted through the noisy channel and the decoded vector is denoted as $\widehat{\boldsymbol L}$. Then, $\widehat{\boldsymbol L}$ is fed into the semantic decoder to obtain the corresponding semantic task output, which in this case is the classification result.

We aim to design the channel encoder, channel decoder, and semantic decoder to achieve efficient and robust point cloud classification transmission.

\begin{figure*}[tbp]
\includegraphics[width=0.69\textwidth]{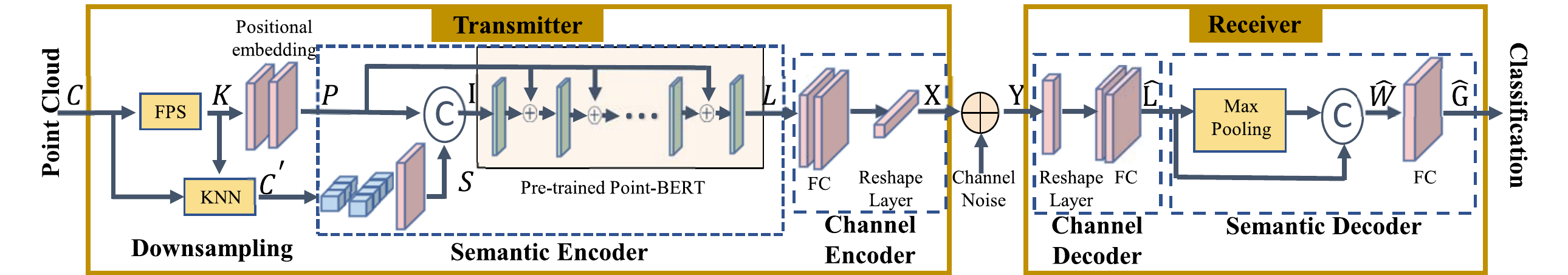} 
\centering % central position
\caption{The overall architecture of the proposed point cloud classifications semantic communication system}
\label{model}
\end{figure*}

\subsection{Performance Metric}
The performance metric of the proposed semantic communication framework for point cloud data transmission is classification accuracy, which is defined as the percentage of correctly classified point cloud subsets in the test set. The higher the classification accuracy is, the better the performance of the system in classifying point cloud data.

\section{Proposed Method}

Our proposed method is a task-oriented semantic communication system aiming to efficiently and robustly classify point cloud data. As can be seen in Fig.~2, our approach consists of four main components: Downsapmling, Semantic Encoder, Channel Encoder and Decoder, and Semantic Decoder. The Downsapmling is combined with FPS downsampling and kNN grouping. Additionally, we use a two-stage training method for uniform adoption in different channel conditions.

\subsection{Downsampling}

The Semantic Encoder is responsible for extracting semantic information from the input point cloud data. Initially, the input dense point cloud $\boldsymbol C \in\mathfrak R^{B\times 1024 \times 3}$ undergoes downsampling using FPS to obtain 64 key positional points $\boldsymbol K \in\mathfrak R^{B\times 64 \times 3}$, where $B$ is the batch size, 1024 is the number of dense points, and the three channels correspond to x, y, and z, respectively. FPS effectively reduces noise and computational complexity while preserving crucial geometric details. The algorithm operates as follows:

\begin{algorithm}[htb] \caption{Farthest Point Sampling Algorithm} \label{pretraining algorithm} \begin{algorithmic}[1]

    \STATE Randomly select a point from the point cloud as the first sample point.
    \STATE Calculate the distance from the remaining points to the first sample point and select the farthest point as the second sample point.

    \STATE \textbf{Repeat:} Repeat step 2 until the number of sample points reaches a predefined threshold.
\end{algorithmic}
\end{algorithm}

After downsampling, we obtain 64 key geometric points, which alone are insufficient for point cloud classification tasks. To enrich the semantic information, we use a kNN algorithm to partition the input points into subsets $\boldsymbol C^{'} \in\mathfrak R^{B\times 64 \times 32 \times 3}$ that are close to the 64 key points, with each sub-cloud containing 32 points. Meanwhile, the key geometric points $\boldsymbol K$ are fed into two FC layers, called learned positional layers, to transform points into positional embeddings $\boldsymbol P \in\mathfrak R^{B\times 64 \times 384}$ for further processing. 

\subsection{Semantic Encoder}

Subsequently, these sub-clouds $\boldsymbol C^{'}$ are fed into two successive one-dimensional convolutions and one fully connected layer (FC), which are lightweight and efficient neural networks, to extract semantic information $\boldsymbol S \in\mathfrak R^{B\times 64 \times 384}$ of the sub-cloud. Then, $\boldsymbol P $ and $\boldsymbol S$ are concatenated together with a learned $ {CLS\_Token\_Embedding} \in\mathfrak R^{B\times 1 \times 384}$ and $ {CLS\_Positional\_Embedding} \in\mathfrak R^{B\times 1 \times 384}$ to get $\boldsymbol I \in\mathfrak R^{B\times 65 \times 768}$. This information is then passed through the pre-trained transformer and results in $\boldsymbol L \in\mathfrak R^{B\times 65 \times 384}$, which is trained by Point-BERT. Point-BERT is a self-supervised masked point model, adopted from the famous masked language model, BERT. The self-supervised masked point model can leverage the intrinsic semantic information of data, which will be explained carefully in the following part.

\subsection{Pre-trained Point-BERT: A Self-supervised Masked Point Model}

The pre-trained Point-BERT model plays a crucial role in our proposed system due to its ability to efficiently extract high quality semantic information from point cloud data. Point-BERT is based on the transformer architecture, which leverages self-attention mechanisms and pre-training strategies to capture complex relationships between points in point cloud data. By utilizing a pre-trained Point-BERT model, we can benefit from transfer learning to adapt the model to specific classification tasks with minimal additional training.

The key feature of Point-BERT is its self-supervised pre-training strategy, inspired by BERT, which eliminates the need for labeled data during training. This characteristic allows the model to be used for a diverse range of 3D point cloud analysis tasks, especially when labeled data is limited or unavailable.

The pre-training process of Point-BERT consists of several steps:

\begin{enumerate}
\item Train dVAE (Discrete Variance Auto-Encoder) to obtain the point tokenizer: Use DGCNN as the tokenizer network within dVAE to map point embeddings into discrete point tokens, which correspond to local geometric patterns. Train the dVAE with Gumbel-softmax relaxation and a uniform prior to handle discrete latent point tokens.

\item Sub-cloud generation and tokenization: Generate sub-clouds by partitioning the input point clouds into smaller, overlapping regions. Apply the trained dVAE tokenizer to convert the point cloud data into discrete point tokens, suitable for processing by a transformer-based model.

\item Learn relationships and masked point modeling with a BERT-style pre-training model: Train a transformer-based model on the tokenized sub-clouds to capture complex relationships and structures. Use a masked point modeling technique by masking a percentage of points in the input data, and train the model to predict the masked points' attributes, enhancing its representation learning capabilities.
\end{enumerate}

By leveraging self-attention mechanisms, pre-training strategies, and transfer learning, Point-BERT captures the underlying semantics of point cloud data, enabling more accurate classification and understanding.

\subsection{Channel Encoder and Decoder}

The semantic embeddings $\boldsymbol L$ obtained from the pre-trained transformer are passed through the Channel Encoder, which consists of two fully connected layers, and results in $\boldsymbol X \in\mathfrak R^{B\times 780 \times 2}$. The encoded data $\boldsymbol X$ is then transmitted over a noisy communication channel. At the receiver end, the noisy data $\boldsymbol Y \in\mathfrak R^{B\times 780 \times 2}$ is processed by the Channel Decoder, which also employs two fully connected layers. The decoder corrects errors introduced by the channel noise and reconstructs the semantic embeddings $\widehat {\boldsymbol L}$.

\subsection{Semantic Decoder}

At the receiver end, the received semantic embeddings $\widehat {\boldsymbol L}$ are utilized to perform classification. A new embedding consists of $\widehat {\boldsymbol W} $, the classification embedding, and max-pooling of semantic embeddings from $\widehat {\boldsymbol L} \in\mathfrak R^{B\times 768}$. The embeddings are subsequently fed into a task-specific classification network comprising a fully connected layer and result in the predicted classification $\widehat {\boldsymbol G} \in\mathfrak R^{B\times 40}$, where 40 represents the number of object categories in the dataset.

\subsection{Two-Stage Training}

In our approach, we employ a two-stage training strategy to ensure that the system is both efficient and adaptable to various channel conditions. 
In the first stage, we train the semantic encoder and semantic decoder, and ignore the channel encoder and decoder by directly inputting the output of the semantic encoder to the semantic decoder, and we use the pretrained Point-BERT from \cite{point-bert}. We use a task-specific classification cross-entropy loss function to optimize the weights of the network.

In the second stage, we train the entire semantic communication system, including the Channel Encoder and Decoder, with a task-specific classification loss function and the mean squared error (MSE) between the semantic embeddings before the Channel Encoder and after the Channel Decoder. During this stage, the pre-trained transformer and positional embeddings from the first stage are frozen, ensuring that the learning achieved in the first stage is preserved.

During the second stage, we introduce a unique aspect to our training process that sets our method apart from other works. Instead of training the network separately for each channel condition, we randomize the signal-to-noise ratio (SNR) between 0 dB and 20 dB for each epoch. This allows our trained network to adapt to different SNR conditions, ultimately saving valuable training time and storage resources. By randomizing the SNR during training, our model can generalize to various channel conditions without needing to retrain for specific SNR values or store separate model weights.

\begin{table}[]
\centering
\footnotesize
\caption{Parameter settings of the proposed semantic communication Network for point cloud classification}
\label{table 2}
\begin{tabular}{|c|c|c|c|}
\hline
                                                                                          & Layer Name               & Parameters           & Activation            \\ \hline
\multirow{3}{*}{\begin{tabular}[c]{@{}c@{}}Learned Positio-\\ nal Embedding\end{tabular}} & FC                       & 128                  & GELU                  \\ \cline{2-4} 
                                                                                          & \multirow{2}{*}{FC}      & \multirow{2}{*}{384} & \multirow{2}{*}{NONE} \\
                                                                                          &                          &                      &                       \\ \hline
\multirow{11}{*}{\begin{tabular}[c]{@{}c@{}}Semantic\\ Encoder\end{tabular}}              & Conv1d                   & 3/128/1              & NONE                  \\ \cline{2-4} 
                                                                                          & BatchNorm1d              & 128                  & ReLU                  \\ \cline{2-4} 
                                                                                          & Conv1d                   & 128/256/1            & NONE                  \\ \cline{2-4} 
                                                                                          & Maxpool                  &             NONE         & NONE                  \\ \cline{2-4} 
                                                                                          & Conv1d                   & 512/512/1            & NONE                  \\ \cline{2-4} 
                                                                                          & BatchNorm1d              & 512                  & ReLU                  \\ \cline{2-4} 
                                                                                          & Conv1d                   & 512/256/1            & NONE                  \\ \cline{2-4} 
                                                                                          & FC                       & 384                  & NONE                  \\ \cline{2-4} 
                                                                                          & CLS\_Token\_Embedding    & 384                  & NONE                  \\ \cline{2-4} 
                                                                                          & CLS\_Position\_Embedding & 384                  & NONE                  \\ \cline{2-4} 
                                                                                          & 12$\times$Transformer           & 6 heads              & NONE                  \\ \hline
\multirow{2}{*}{\begin{tabular}[c]{@{}c@{}}Channel\\ Encoder\end{tabular}}                & FC                       & 512                  & ReLU                  \\ \cline{2-4} 
                                                                                          & FC                       & 24                   & NONE                  \\ \hline
\multirow{2}{*}{\begin{tabular}[c]{@{}c@{}}Channel\\ Decoder\end{tabular}}                & FC                       & 512                  & ReLU                  \\ \cline{2-4} 
                                                                                          & FC                       & 284                  & NONE                  \\ \hline
\multirow{2}{*}{\begin{tabular}[c]{@{}c@{}}Semantic\\ Decoder\end{tabular}}               & \multirow{2}{*}{FC}      & \multirow{2}{*}{40}  & \multirow{2}{*}{NONE} \\
                                                                                          &                          &                      &                       \\ \hline
\end{tabular}
\end{table}

\section{Experimental Results}

In this section, we present the experimental results of our proposed method on the ModelNet40 dataset and compare our method with state-of-the-art baselines. We analyze the computational complexity and speed, and demonstrate the impact of various components of our proposed system through ablation studies.

\subsection{Experimental Setup}

In this study, we assess the performance of our proposed method by comparing it to other deep learning-based semantic communication systems designed for point cloud classification tasks. To carry out the evaluation, we consider the AWGN channel. The ModelNet40 dataset is utilized for both training and testing, as it is a widely-recognized benchmark in the field of point cloud classification. The dataset comprises 9,843 point clouds in the training set and 2,468 point clouds in the test set.

We adopt the existing semantic communication approach by \cite{Branchy}, known as Branchy-GNN, as our benchmark for comparison. Our Baseline is determined by employing the proposed semantic encoder and decoder, while disregarding the noisy channel and the channel encoder and decoder. This provides the upper-bound performance. The detailed configuration of our proposed network is presented in Table \ref{table 2}. All three methods, including our proposed approach, are trained and tested using the same datasets.
The hyperparameters applied in this study include:
\begin{itemize}
\item Batch size: 32
\item Optimizer: AdamW optimizer
\item Learning rate scheduler: SGDR
\item Learning rate decay: 0.1
\end{itemize}

During the training phase, we establish channel conditions for our method with SNR values varying randomly from 0 dB to 20 dB. Conversely, we train and test the Branchy-GNN approach under identical SNR conditions, adhering to the methodology outlined in \cite{Branchy}.

\subsection{Results under Different Channel Conditions}
\begin{figure}[htbp]
\centering
\includegraphics[width=\linewidth]{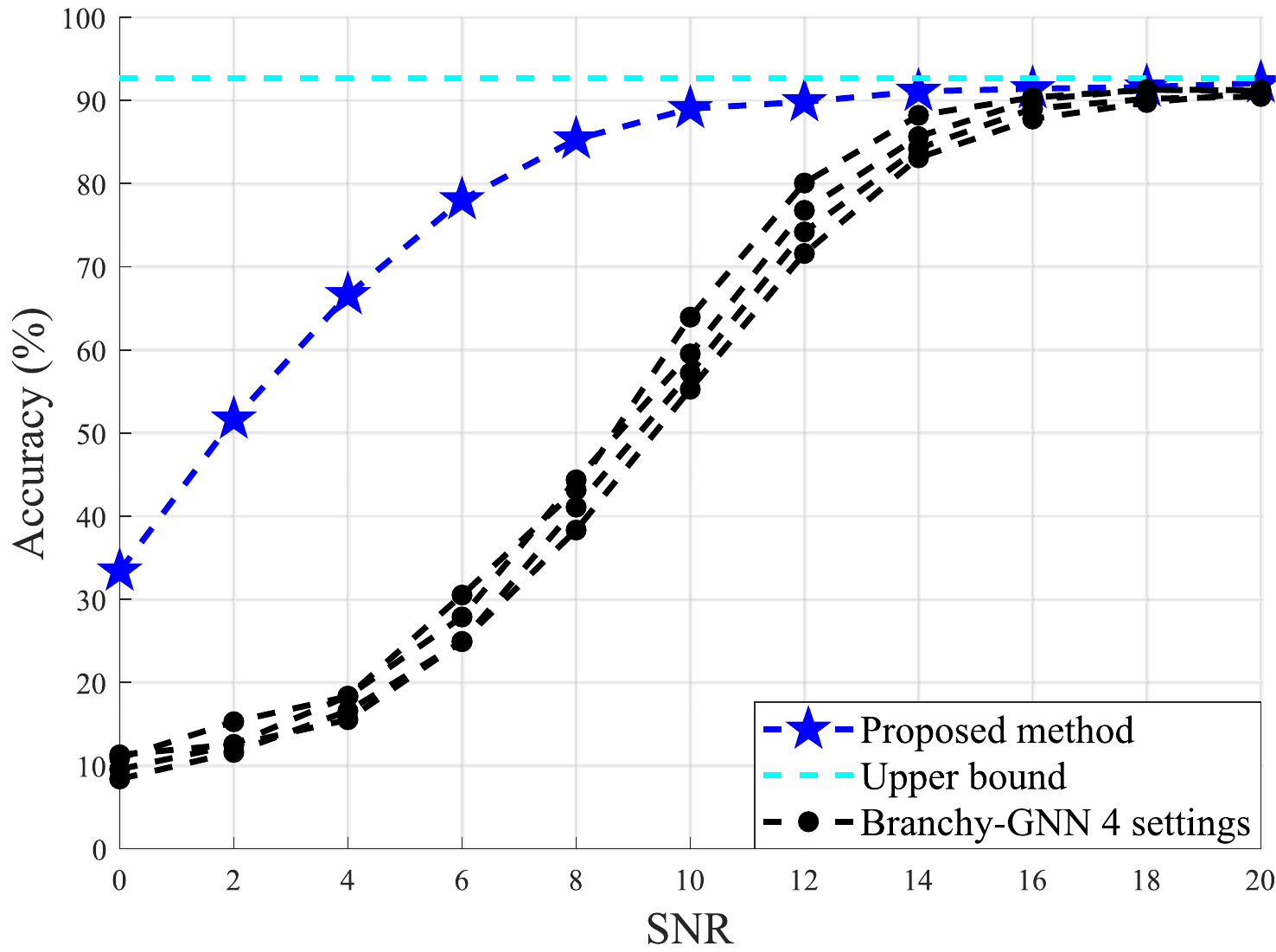}
\caption{Point Cloud classification accuracy under different channel conditions.The proposed method achieves significantly superior classification accuracy (above 78\% at 6 dB SNR) compared to Branchy-GNN (24.9-30.5\% at 6 dB SNR) across varying SNRs from 0 dB to 20 dB.}
\label{fig:results_channel_conditions}
\end{figure}

\begin{figure}[htbp]
\centering
\includegraphics[width=\linewidth]{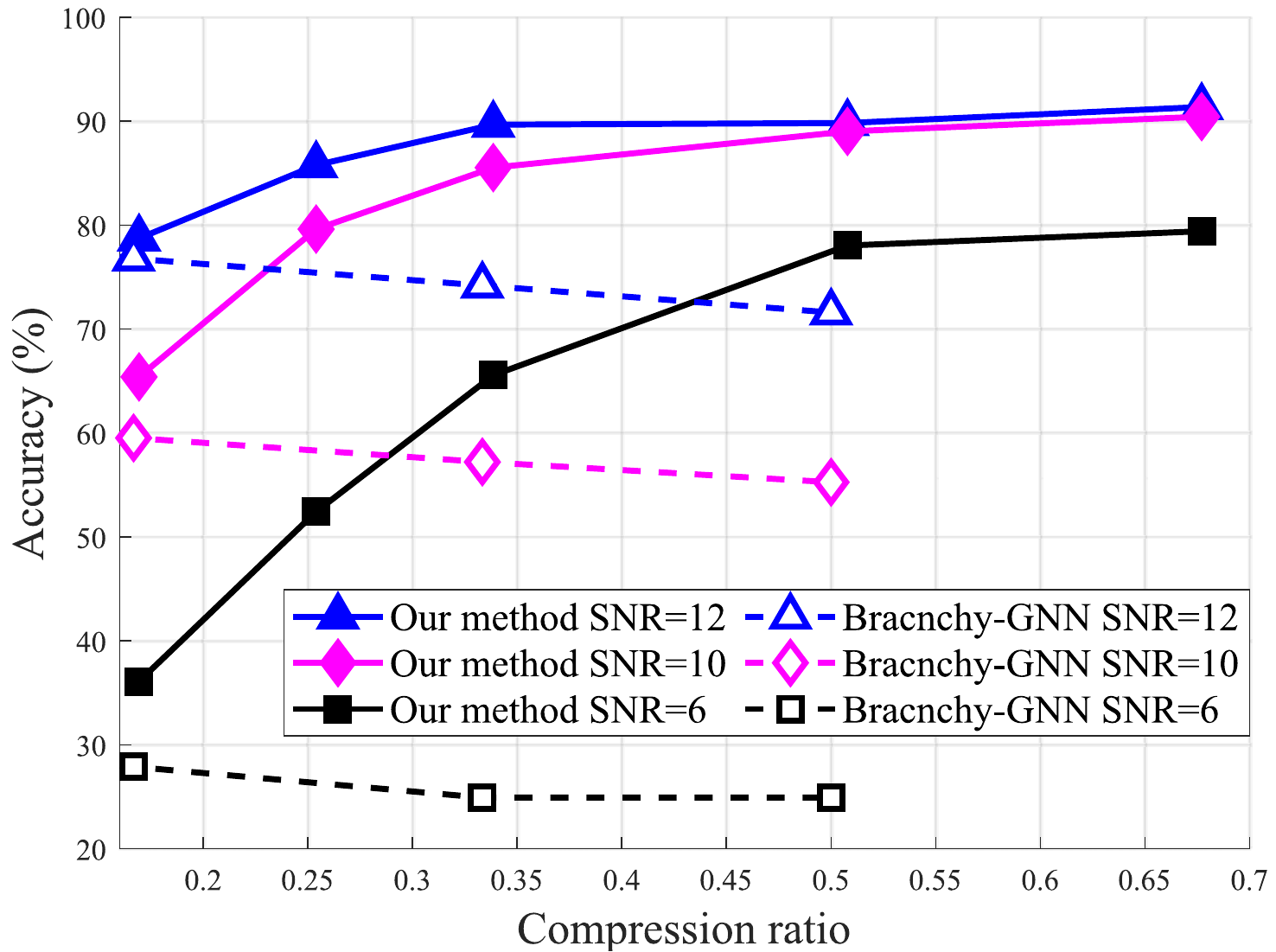}
\caption{Point Cloud classification accuracy under different compression efficiency and channel conditions. With nearly the same compression ratio, the proposed method achieves significantly superior classification accuracy compared to Branchy-GNN across varying SNRs from 0 dB to 20 dB, especially in low SNR regimes, highlighting its effectiveness in challenging communication environments.}
\label{fig:results_compression_ratio}
\end{figure}

In this subsection, we present the results of our proposed point cloud semantic communication system under varying channel conditions. Fig.~ \ref{fig:results_channel_conditions} and Fig.~ \ref{fig:results_compression_ratio} show a comparison of the classification accuracy and compression ratio across different scenarios. To evaluate the performance of our approach, we compared it with the existing Branchy-GNN method in terms of accuracy and communication resource consumption. Specifically, we varied the signal-to-noise ratio (SNR) from 0 dB to 20 dB and measured the point cloud classification accuracy achieved by both methods under the same communication overhead. And we follow \cite{jscc} to define the compression ratio as the ratio of the dimension of vectors ($\boldsymbol C$) to be transmitted and the size of the input dense point cloud ($\boldsymbol X$).

Our experimental results showed that our proposed method outperforms the Branchy-GNN approach under all channel conditions. When the SNR was larger than 10 dB, our method achieved a classification accuracy of over 89\% and consistently outperformed the Branchy-GNN method by at least 0.8\%. This result not only demonstrates the high accuracy of our method but also highlights its superiority in terms of discrimination ability in comparison to the existing method.

Furthermore, as the channel conditions worsened, our proposed method also showed its superior robustness in achieving high accuracy. As the SNR decreased, the classification accuracy of the proposed approach decreased much slower than the Branchy-GNN method. Specifically, while the classification accuracy of the Branchy-GNN method dropped significantly when the SNR was less than 10 dB, our proposed approach remained effective, with an accuracy above 33\%. Even in the harshest conditions, where the SNR was only 6 dB, our proposed method still achieved an accuracy greater than 78\%, 48\% higher than Branchy-GNN. This indicates that our approach can overcome the adverse effects of noisy and unreliable communication channels, providing more reliable semantic communication of 3D information.

To clarify, as shown in Fig.~ \ref{fig:results_compression_ratio}, our proposed method and the Branchy-GNN method used the same amount of communication resources. However, our method demonstrated better stability and reliability under different channel conditions, leading to higher accuracy.

In summary, our proposed point cloud semantic communication system demonstrates strong robustness and achieves state-of-the-art classification accuracy under various channel conditions, especially in harsh environments. It shows much better performance than existing methods, and it's also resource-efficient. These results suggest that the proposed system can be a promising solution to enable reliable, and efficient 3D point cloud semantic communication.

\subsection{Computational Complexity and Speed}

In this subsection, we discuss the computational complexity and speed of our proposed point cloud semantic communication system, as well as compare it with the existing Branchy-GNN method. The comparison is based on a table that contains information on the time taken to transmit all the dataset's test subset, average accuracy (Acc-avg), accuracy at SNR=20 dB (Acc snr=20), and accuracy at SNR=8 dB (Accu snr=8) for both our method and the Branchy-GNN method.

From the table \ref{Complexity}, we can see that our proposed method takes 0.1166 seconds to process all the test data, with an average accuracy of 78.2\%, 92.1\% accuracy at SNR=20 dB, and 85.3\% accuracy at SNR=8 dB. On the other hand, the Branchy-GNN method has different processing times and accuracies depending on the number of branches used (Branchy-GNN\_1 to Branchy-GNN\_4), which are all worse than ours.

It is evident that our proposed method is not only faster than the Branchy-GNN\_4 but also achieves higher average accuracy across different SNR conditions. While the Branchy-GNN\_1 method is faster than our proposed method, it comes at the cost of lower accuracy in all SNR conditions.

In summary, our proposed point cloud semantic communication system shows a good balance between computational complexity and speed, as well as classification accuracy in different SNR conditions. This makes it a promising solution for reliable and efficient 3D point cloud semantic communication, particularly when compared with existing methods like the Branchy-GNN.

\begin{table}[]
\caption{Comparison of computational complexity, speed, and classification accuracy between the proposed method and Branchy-GNN under different SNR conditions.}
\begin{tabular}{|c|c|c|c|c|}
\hline
               & time(s) & Acc-avg & Acc snr=20 & Acc snr =8 \\ \hline
Proposed       & 0.1166  & 78.2\%             & 92.1\%            & 85.3\%            \\ \hline
Branchy-GNN\_4 & 0.1575  & 56.7\%           & 91.2\%            & 43.2\%            \\ \hline
Branchy-GNN\_3 & 0.0759  & 55.4\%             & 91.2\%            & 44.4\%            \\ \hline
Branchy-GNN\_2 & 0.0452  & 53.4\%             & 90.3\%            & 41.1\%            \\ \hline
Branchy-GNN\_1 & 0.0351  & 52.7\%             & 91.0\%            & 38.3\%            \\ \hline
\end{tabular}
\label{Complexity}
\end{table}

\subsection{Ablation Studies}

In this subsection, we conduct ablation studies to further evaluate the effectiveness of our proposed method and highlight its advantages. The primary aim of these studies is to investigate the performance of our approach without the use of a pre-trained model and with the application of our proposed two-stage training.

\begin{figure}[htbp]
\centering
\includegraphics[width=\linewidth]{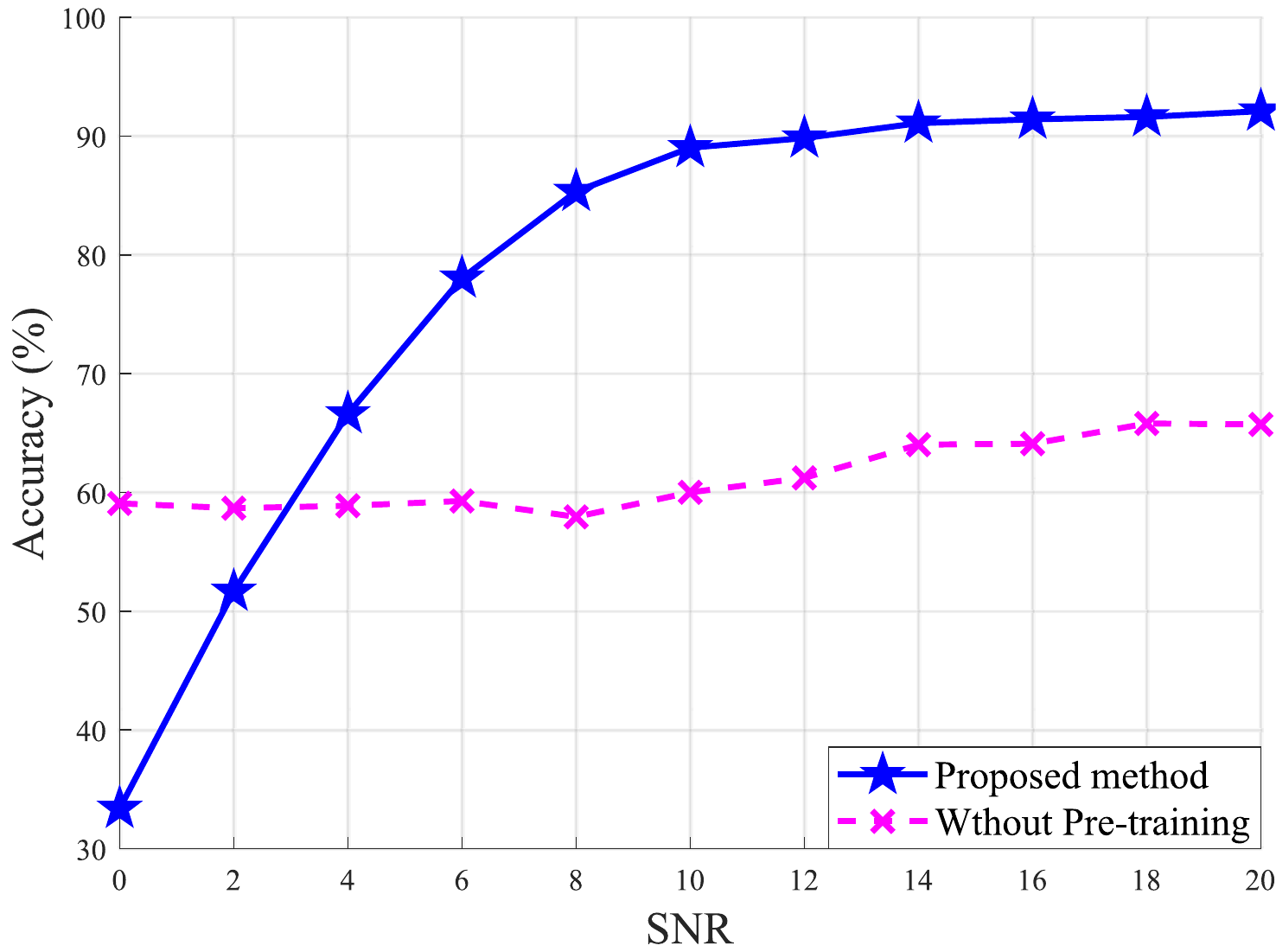}
\caption{Results of ablation studies for different components of the proposed method. Using a pre-trained model and two-stage training process significantly improves the overall classification accuracy of the proposed semantic communication system across various SNR conditions.}
\label{fig:ablation_studies}
\end{figure}

To conduct the ablation study, we performed a simple experiment comparing the classification accuracy of point clouds under varying signal-to-noise ratio (SNR) conditions, ranging from 0 dB to 20 dB, as shown in Fig. \ref{fig:ablation_studies}. In the first scenario, the classification accuracy of our proposed method was around 60\%, which is significantly lower than the accuracy achieved when using the pre-trained model and two-stage training process. On the other hand, when applying the pre-trained model and our proposed two-stage training method, the classification accuracy improved significantly.

These results highlight the superiority of our proposed method in achieving higher overall classification accuracy. It is worth mentioning that without using a pre-trained model, the system learns strong channel noise characteristics at the cost of point cloud semantic knowledge. Although the accuracy at an SNR of 0 dB is 59.08\%, the overall accuracy remains low, and the accuracy at higher SNR values is only 65.72\%. This is likely because the model focuses only on learning to handle channel noise rather than retaining sufficient understanding of point cloud semantics.

In summary, the ablation studies demonstrate the effectiveness and advantages of our proposed point cloud semantic communication system employing a pre-trained model and a two-stage training process. The results indicate that our method can achieve significantly higher classification accuracy across various SNR conditions, outperforming the approach without pre-training. This suggests that the integration of a pre-trained model and two-stage training is essential to improving the performance of point cloud classification in challenging communication environments.

\section{Conclusion}

In this paper, we have presented a task-oriented semantic communication system for effective and robust point cloud transmission. Our approach leverages FPS and kNN to partition the point cloud into sub-clouds, which are then processed by a pre-trained semantic encoder to generate a task-relevant feature vector for transmission via the physical channel. At the receiver end, a semantic decoder performs the classification task based on the recovered feature vector. Simulation results validate the effectiveness and robustness of the proposed system.

The proposed method addresses the challenges faced by existing point cloud communication systems, such as high communication overhead and susceptibility to noise, by incorporating a task-relevant feature extraction mechanism and a two-stage training process. Moreover, our approach outperforms previous solutions in terms of classification accuracy across SNR conditions from 0 dB to 20 dB, highlighting the benefits of our method in improving point cloud classification performance in challenging communication environments.

Our method demonstrates a balance between accuracy and speed, making it a promising solution for reliable and efficient 3D point cloud semantic communication. By employing a pre-trained model and a two-stage training process, our approach efficiently handles noisy channel interference, maintaining high classification accuracy even under low SNR conditions from 0 dB to 6 dB..

In conclusion, this work makes important contributions to the development of efficient and robust point cloud communication systems, paving the way for future research in semantic communication and task-oriented transmission systems. Future work could explore other point cloud processing tasks, such as segmentation or registration, to broaden the applicability of the proposed semantic communication framework.

\vspace{12pt}

\end{document}